\documentclass[twocolumn,showpacs,preprintnumbers,amsmath,amssymb]{revtex4}

\usepackage{graphicx}
\usepackage{dcolumn}
\usepackage{bm}
\usepackage{amssymb} 
\usepackage[usenames]{color}

\begin{document}

\title{High Resolution Study of Magnetic Ordering at Absolute Zero}

\author{M. Lee$^{1}$, A. Husmann$^{2}$, T.F. Rosenbaum$^{1}$,  and G. Aeppli$^{3}$}

\affiliation{$^{1}$James Franck Institute and Department of Physics,
 University of Chicago, Chicago, Illinois 60637\\
 $^{2}$Toshiba Research Europe Ltd., 260 Cambridge Science Park, Cambridge, CB4 0WE, UK\\
$^{3}$London Centre for Nanotechnology and Department of Physics and Astronomy, University College London, London, WC1E 6BT, UK}

\date{\today}

\begin{abstract}
    
High fidelity pressure measurements in the zero temperature limit provide a unique opportunity to study the behavior of strongly interacting, itinerant electrons with coupled spin and charge degrees of freedom. Approaching the exactitude that has become the hallmark of experiments on classical critical phenomena, we characterize the quantum critical behavior of the model, elemental antiferromagnet chromium, lightly doped with vanadium.  We resolve the sharp doubling of the Hall coefficient at the quantum critical point and trace the dominating effects of quantum fluctuations up to surprisingly high temperatures. 

\end{abstract}

\maketitle
Phase transitions at the absolute zero of temperature are a result of Heisenberg's uncertainty principle rather than of a thermal exploration of states. Their ubiquity in materials of large technological interest, including transition metal oxides and sulfides, metal hydrides, superconducting cuprates, and colossal magnetoresistance manganites, combined with the intellectual challenge presented by many strongly interacting quantum degrees of freedom, places quantum phase transitions at the core of modern condensed matter physics. Quantum fluctuations inextricably intertwine the static and dynamical response of the material changing state, introducing new critical exponents, new scaling laws, and new relationships between the spin and charge degrees of freedom \cite{Sachdev, Sondhi, Chakravarty89}.

Varying the quantum fluctuations required for zero-temperature phase transitions is more difficult than changing temperature, with the result that the understanding of quantum phase transitions is far less detailed than that of their classical analogues. Typically, fluctuations are varied by scanning the composition of an alloy, such as the doped cuprates that host high-temperature superconductivity. The result is that a new sample, each with unique disorder, must be fabricated for every zero-point fluctuation rate sampled.  Tuning the transition with an external magnetic field, a quantity that is easily and precisely regulated, is a cleaner technique and provides correspondingly greater detail for both insulators \cite{Bitko} and metals \cite{SRO}. However, magnetic fields also break time-reversal symmetry, which is particularly significant for quantum phase transitions because the dynamics responsible for zero-point fluctuations are altered profoundly. It is important, therefore, to examine a quantum phase transition for a simple material with high precision without applying a symmetry-breaking field. In response to this challenge, we have performed a high-resolution hydrostatic pressure study of a model quantum phase transition: elemental chromium diluted with its neighbor vanadium, small amounts of which can smoothly suppress Cr's spin-density-wave transition to $T = 0$. The Cr-V single crystals permit tuning with high fidelity and we are able to characterize precisely the signatures of vanishing magnetic order in a system sufficiently simple to promise theoretical tractability. 

Cr is the archetypical metallic antiferromagnet for which conduction electrons are lost as they order magnetically when the temperature passes below the N\'{e}el temperature $T_{N}$. The loss of carriers is most dramatically seen in the Hall effect, which measures the density of free, metallic carriers. The fundamental issue at a metallic quantum critical point (QCP) is what occurs to the free carriers that eventually become localized due to magnetic order: should they be counted as free or as localized? In a previous experiment, we have examined alloys of Cr$_{1-x}$V$_{x}$ and tracked the loss of carriers as a function of $x$. We found a jump in the Hall number at the QCP, indicating giant fluctuations in the number of free carriers, even though other measures, such as the N\'{e}el temperature and the internal magnetization vanished continuously \cite {Yeh}. The present work demonstrates that if we resolve the quantum critical point with two orders of magnitude greater precision than is possible by varying alloy composition, the Hall number does eventually undergo a continuous evolution at the $T = 0$ phase transition. Moreover, we are able to fix three critical exponents that characterize this quantum many-body problem and find that the diagonal and off-diagonal elements of the resistivity tensor behave differently through the QCP.

We applied hydrostatic pressure to single crystals of Cr$_{0.968}$V$_{0.032}$ of 1 mm$^{3}$ volume with a BeCu piston-anvil cell with a WC insert. Tuning the $T = 0$ transition with pressure fixes the disorder from the V substitution and, in fact, we find that the $T \rightarrow 0$ disorder scattering (paramagnetic contribution to the longitudinal resistivity) is independent of applied pressure. At $x = 0.032$, the sample has a N\'{e}el temperature $T_{N} = 52 K$, significantly reduced from $T_{N} = 311 K$ for pure Cr and carefully chosen to sit on the leading edge of the jump in the Hall coefficient \cite {Yeh} at $x_{c} \sim 0.034$. The pressure cell was mounted in the bore of a superconducting magnet in a $^{3}$He cryostat to reach sub-Kelvin temperatures. The single crystals of Cr$_{1-x}$V$_{x}$ were grown by Ames Laboratory in an arc zone refining furnace and annealed for 72 hours at 1600$^\circ$C to relieve internal strain.  Lau\'{e} backscattering was used to determine the orientation of the bcc lattice, and several pieces aligned with the (100) crystallographic direction were cut from the center of each boule using spark erosion. Proper etching to remove the surface oxide (with a 3:1::HCl:H2O2 solution quenched with THF) was found to be essential for reproducible results. The long axis of the crystal was mounted in the pressure cell perpendicular to the applied field with Fluorinert as the pressure medium and a chip of (V$_{0.99}$Ti$_{0.01}$)$_{2}$O$_{3}$ serving as the manometer. We could measure relative values of $P$ to better than 0.1 kbar. The diagonal and off-diagonal components of the resistivity were measured using a conventional five-probe ac-bridge technique in the Ohmic and frequency-independent limits. The Hall coefficient was obtained in the linear field regime via sweeps from Ð1.2 T to +1.2 T. 
\begin{figure}
\includegraphics[width= 3.75 in]{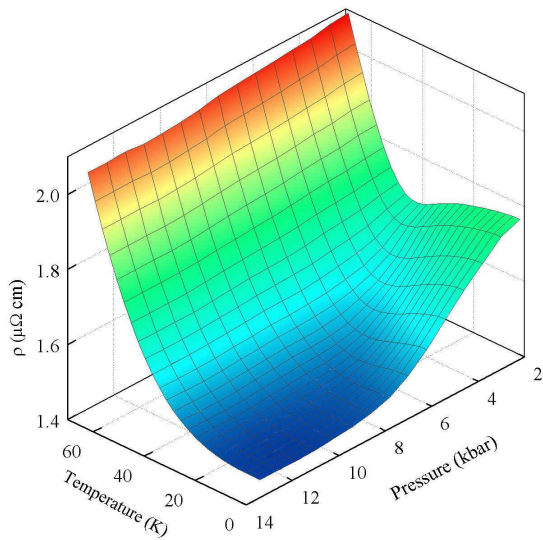}
\caption{\label{Fig:figure1}
Variation of the longitudinal resistivity $\rho$ with temperature and pressure (in descending order) in the immediate vicinity of the quantum critical point for single crystal Cr$_{0.968}$V$_{0.032}$. The minimum in the resistivity tracks the opening of the spin-density-wave gap, which is completely suppressed at the $T = 0$ critical pressure, $P_{c} = 7.5$ kbar.} 
\end{figure}

The variation of the longitudinal resistivity as a function of pressure and temperature in the immediate vicinity of the QCP is captured in Fig. 1. The minimum in $\rho(T)$ at T$_{min}$ marks the opening of the spin-density-wave gap \cite {Alexander_Carter}.  It collapses as a function of pressure, with the low-temperature rise in $\rho (T)$ disappearing at $P_{c} = 7.5$ kbar. The temperature dependence of $\rho$ for $T > T_{min}$ when $P < P_{c}$ and for all $T  < 100$ K for $P  > P_{c}$ follows a $T^{3}$ power law expected from phonon scattering in a disordered metal alloy \cite {Campbell}.  Although $T_{min}(P)$ tracks $P_{c}$, the full T-dependence of the electrical resistivity at $P \sim P_{c}$ appears oblivious to the existence of the QCP, even under the magnifying glass of pressure tuning. This smooth variation through $P_{c}$ contrasts sharply with the Ònon-Fermi liquidÓ behavior seen in some superconducting cuprates \cite {electronbook}, heavy fermion compounds \cite {Stewart} and metamagnets \cite {SRO}, where a reduced power law in T emerges at the QCP. We note that we find no evidence for a coexisting superconducting state near $P_{c}$ \cite {Fawcett94} for T $>$ 0.4 K. 

\begin{figure}
\includegraphics[width= 2.25 in]{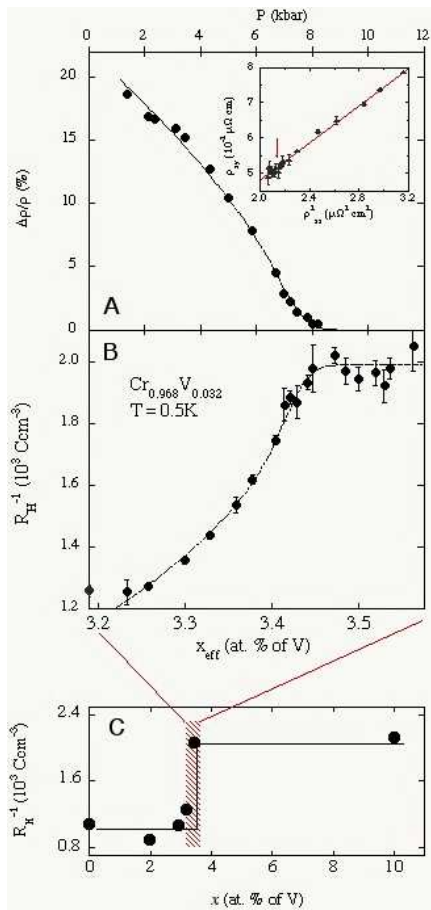}
\caption{\label{Fig:figure2}
Critical behavior of the (a) normalized resistivity and (b) the inverse Hall coefficient in the T$\rightarrow$0 limit as a function of pressure $P$. The resistivity and the inverse Hall coefficient approach the quantum phase transition with different critical exponents.  The curves are fits to $(P-P_{c})^{0.68\pm0.03}$ and $(P-P_c)^{0.50\pm0.02}$, respectively, convolved with a Gaussian spread in stoichiometry (see text).  The exploded view at the bottom illustrates the high resolution afforded by pressure measurements as compared to results [6] from a series of crystals with different vanadium concentrations. The inset in (a) shows that $\rho_{xy} \sim \rho_{xx}^2$, implying that the transverse conductivity $\sigma_{xy}(T=0)$ is actually constant through the QCP. }
\end{figure}

The approach of the T $\rightarrow$ 0 longitudinal resistivity to $P_{c}$ from below is one key measure of the quantum critical behavior. The opening of the spin-density-wave gap leads to an excess, normalized resistivity  
\begin{equation}
\frac{\Delta\rho}{\rho}(T) = \frac{\rho(T) -\rho_{P}}{\rho}
\end{equation}
where the baseline paramagnetic resistivity $\rho_{P}$ is determined from the extrapolation of the $T^{3}$ fit for $T > T_{min}$ to $T = 0$. We focus in Fig. 2a on the zero temperature limit of Eq. (1). $\Delta\rho/\rho$ falls continuously to zero as a function of pressure with a critical exponent less than one.  The rounding in the immediate vicinity of $P_{c}$ can be attributed to small differences in V concentration across the sample. We fit the data to a critical form $\Delta\rho/\rho($T=0.5K, P$) \sim (P_{c}-P)^{\beta}$, convolved with a Gaussian distribution of critical pressures.  The three-parameter fit (solid line in Fig. \ref{Fig:figure2}a) yields a critical exponent $\beta = 0.68 \pm 0.03$ at the critical pressure $P_{c} = 7.5 \pm 0.1$ kbar. This $P_{c}$ corresponds to a mean $x_{0} = 3.189 \pm 0.001\%$, close to the nominal V concentration of $3.2\%$ and set by the empirical formula $P_{c} = 31.042\cdot (3.430-x)$ which is determined from a consideration of all available data reported in the literature (discussed below). The observed tail in $\Delta\rho/\rho(P)$ gives a width $\delta x = 0.019 \pm 0.001\%$.  

Perhaps the most striking feature associated with the QCP is the 100$ \%$ jump in the zero temperature carrier density over a very narrow range of $x$ (Fig. \ref{Fig:figure2}c).  Fixing $x$ and tuning the transition with $P$ permits an investigation of whether this ÒjumpÓ is sharp but continuous, and if so, whether there is an additional critical exponent that characterizes the transition. We plot in Fig. \ref{Fig:figure2}c the inverse Hall coefficient ($R_{H}^{-1}$) at $T = 0.5$ K in the immediate vicinity of $P_{c}$. $R_{H}^{-1}(T\rightarrow0)$ assumes separate and fixed values in the antiferromagnet ($P< 2$ kbar) and in the paramagnet ($P > 8$ kbar), but changes continuously between the two regimes.  We fit the data by fixing $\delta x$ from $\Delta\rho/\rho (P)$ and again convolving a critical form $R_{H}^{-1}(T$=$0.5K,P) \sim (P_{c}-P)^{\alpha}$ with a Gaussian distribution of critical pressures in a two-parameter fit (solid line in Fig. \ref{Fig:figure2}b). We find a consistent $P_{c} = 7.5 \pm 0.1$ kbar, but a second critical exponent $\alpha = 0.50 \pm 0.02$. The finite width in $R_{H}^{-1}(P)$ cannot arise from the convolution of a step function (representing a discontinuous transition) with large inhomogeneities in x because: (i) the fit would require an inconsistent, far larger $\delta x$ and (ii) the value of $P_{c}$ would be fixed at such a low pressure that it would be unphysical. 

\begin{figure}
\includegraphics[width=3.25  in]{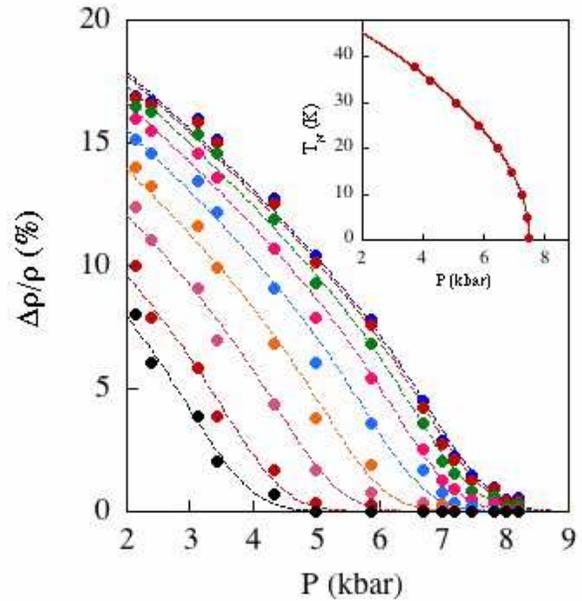}
\caption{\label{Fig:figure3} 
The normalized resistivity can be fit with a critical exponent $2/3$ at all temperatures from 0.5 to 38 K.  Pressure curves shown are for 0.5, 5, 10, 15, 20, 25, 30, 35, and 38 K. In all cases, we fix the spread in stoichiometry to the T = 0 result, $\delta x = 0.019\% $.  Inset: The N\'{e}el temperature approaches zero as $(P-P_c)^{0.49\pm0.02}$.}
\end{figure} 	

The connection of the $T=0$ critical behavior to the finite temperature response is summarized in Fig. \ref{Fig:figure3}. We fix temperature, vary $P$ and find that $\Delta\rho/\rho (P)$ for all $T$ up to 38 K, the highest temperature where we have sufficient data to draw meaningful conclusions, can be fit (dotted lines) with the same critical exponent ($\beta=2/3$) and the identical constant of proportionality $(0.056 \pm 0.003)$. The mean field result, $\beta=1/2$, is not recovered even for P $ \ll $ Pc, indicating the predominant effects of quantum fluctuations up to $T \sim T_{N}$. Strongly enhanced fluctuations have been identified up to 0.5 eV in inelastic neutron scattering studies of Cr$_{0.95}$V$_{0.05}$ \cite {Hayden} and may account as well for the unusual temperature dependence of the Hall resistivity for $T  > T_{N}$ found in Cr-V alloys \cite {Yeh} and familiar from studies of the cuprate superconductors \cite {Ong}. Finally, we can use the $P_{c}(T)$ from the fits in the main part of Fig. \ref{Fig:figure3} to construct the $T-P$ phase diagram for Cr$_{0.968}$V$_{0.032}$. The N\'{e}el temperature is suppressed with pressure as $T_{N} \sim (P_{c}-P)^{\gamma}$ with $\gamma = 0.49\pm 0.02$ and $P_{c} = 7.5 \pm 0.1 $ kbar (Fig. \ref{Fig:figure3}, inset).  Isothermal rather than isobaric cuts of the data are essential to determine $\gamma$ accurately given the almost vertical approach of the phase boundary to the pressure axis as $T_{N} \rightarrow 0$.

\begin{figure}
\includegraphics[width=3.25 in]{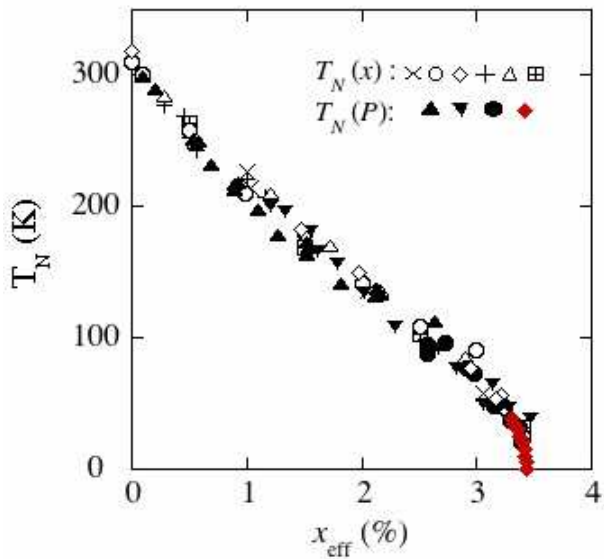}
\caption
{\label{Fig:figure4} Collapse onto a universal curve of all available data on the pressure and doping dependence of the N\'{e}el temperature in Cr$_{1-x}$V$_{x}$, assuming a linear conversion between $P$ and $x$.  Data very close to the QCP are required to reveal the true critical behavior. $T_{N}$ from various $x$'s: $\times$ \cite {Arajs}, $\circ$ \cite {Trego}, $\lozenge$ \cite {Yeh},  $+$ \cite {Komura}, $\triangle$ \cite {Barnes},  $\boxplus$ \cite {White} ; from $P$ measurements: pure Cr's $\blacktriangle$ \cite {McWhan}, V 1.2\% $\blacktriangledown$ \cite {Rice69}, V 2.8\% $\bullet$ \cite {Rice69}, V 3.2\% {\color {BrickRed} $\blacklozenge$} (our data).}
\end{figure}

Close proximity to the QCP is an additional requirement for an accurate determination of $\gamma$. We collect in Fig. \ref{Fig:figure4} the variation of $T_{N}$ with $x$ for a broad range of Cr-V alloys measured via electrical transport \cite {Yeh,Arajs,Trego}, neutron diffraction \cite{Komura}, nuclear magnetic resonance \cite {Barnes}, and thermal expansion \cite{White}, as well as the suppression of spin-density-wave order with pressure for pure Cr \cite{McWhan}, Cr$_{0.988}$V$_{0.012}$ \cite{Rice69}, Cr$_{0.972}$V$_{0.028}$ \cite{Rice69} and Cr$_{0.968}$V$_{0.032}$ (our data). All the data can be collapsed onto a universal curve using an effective V concentration $x_{eff}$ that assumes the simplest linear conversion between chemical doping and applied pressure: $P_{c} = 31.042\cdot(3.430-x)$. By the congruence of the data for pure Cr and its alloys, it appears that disorder is not a dominant factor. $T_{N}$ decreases linearly with $x_{eff}$ across almost the entire composition range, only assuming its critical form with exponent $\gamma = 1/2$ very close to $x_{c} \sim 3.430 \%$.  The initial linear onset mimics the suppression of $T_{N}$ with $x$ and $P$ in the heavy fermion antiferromagnet CeCu$_{6-x}$Au$_{x}$\cite{Bogenberger}, where measurements approached $\delta x_{eff}/x_{eff} \sim 5 \times 10^{-3}$.  For the Cr$_{1-x}$V$_{x}$ antiferromagnet, the curvature close to the QCP becomes apparent only because of our pressure experimentÕs resolving power at the critical point, $\delta x_{eff}/x_{eff} \sim 1 \times 10^{-4}$ \cite{Tom}. Although Cr$_{0.968}$V$_{0.032}$ appears to behave critically over almost its entire range of $x_{eff}$, $T_N(P=0) = 52$ K is far reduced from that of pure Cr (311K), which in turn is far below the bare magnetic coupling $\sim$ 0.5 eV.

The two-band itinerant model for antiferromagnetism \cite {Rice69, Fedders} has been used to consider the pressure dependent N\'{e}el temperature of pure Cr and its alloys. It is mathematically identical to the BCS model for superconductivity and predicts $\gamma = 1/2$. A scaling model \cite {Hertz_Milis} for a $z =  2$ antiferromagnetic QCP predicts $\gamma = z/(d-2+z) = 2/3$ in dimension $d = 3$, but does not apply strictly to a magnet with a partially-nested Fermi surface. Recent calculations \cite {Chakravarty02, Norman} consider explicitly the disappearance of itinerant spin density wave order at a QCP and predict confluent critical exponents for $R_{H}^{-1}$ and $T_{N}$. Although we do find $\alpha = \gamma = 1/2$, there is no theoretical basis for an additional exponent to characterize the critical behavior of $\Delta\rho/\rho$:  the experimentally observed $\beta = 2/3$.  If the nesting is imperfect, then different regions of the Fermi surface, Òhot spotsÓ with low curvature and Òcold spotsÓ with high curvature, could dominate different aspects of the critical behavior, akin to theories for high-$T_c$ and heavy fermion superconductors \cite {Hlubina}.

At $T = 0$, the ordinary Hall number is an intrinsic, non-zero quantity, but the longitudinal resistivity Ð which is characterized by the anomalous power law of $2/3$ Ð is non-zero only by virtue of scattering from impurities and other defects. Could the resistivity be tracking a more fundamental quantity, such as the transverse conductivity $\sigma_{xy} = \rho_{xy}/\rho_{xx}^{2}$? We find that $\rho_{xy} \sim \rho_{xx}^{2}$ (Fig. 2 inset), implying that $\sigma_{xy}(T=0)$ is a constant, insensitive to P even as it crosses $P_{c}$. This result depends crucially on tuning the transition with P because varying the composition x alters the disorder and the scattering potential for each sample, precluding any simple relationship between $\rho_{xy}$ and $\rho_{xx}$. A theory of our $\rho_{xx}$ data near $P_{c}$ in Cr$_{0.968}$V$_{0.032}$ can seek to explain either the apparently anomalous power law describing $\Delta \rho/\rho$ or the seeming pressure independence of $\sigma_{xy}$.  
The fact that the asymptotic critical behavior manifests itself only very close to the quantum phase transition emphasizes the need for similar high precision work on other systems, as well as underlining the possibility that much of the behavior of Cr$_{1-x}$V$_{x}$ and Cr itself may be controlled by a more radical type of quantum criticality, namely one where the Fermi surface undergoes sudden collapse \cite {Schroeder}. The simplicity of the Cr-V system provides the means to test directly our fundamental notions of physics at a QCP, including the use of the Hall coefficient as a sensitive diagnostic of the underlying order \cite {Yeh}, and the manifestation of these ideas in real-world materials.

We are grateful to P. Coleman, A.J. Millis and M. Norman for illuminating discussions.  The work at the University of Chicago was supported by the National Science Foundation under Grant DMR-0114798 and that at UCL by a Wolfson Royal Society Research Merit Award.


\begin{thebibliography}{99}

\bibitem{Sachdev} S. Sachdev, {\em Quantum Phase Transitions} (Cambridge University Press, New York, 1999). 
 
\bibitem{Sondhi} S. L. Sondhi {\it et al.}, Rev. Mod. Phys. {\bf 69}, 315 (1997).

\bibitem{Chakravarty89} S. Chakravarty, B.I. Halperin and D.R. Nelson, Phys. Rev. B {\bf 39}, 2344 (1989). 

\bibitem{Bitko} D. Bitko, T.F. Rosenbaum, G. Aeppli, Phys. Rev. Lett. {\bf 77}, 940 (1996).

\bibitem{SRO} S. A. Grigera {\it et al.}, Science {\bf 294}, 329 (2001); G. Aeppli, Y.-A. Soh, Science {\bf 294}, 315 (2001).

\bibitem{Yeh}A. Yeh {\it et al.},  Nature {\bf 419}, 459 (2002).

\bibitem{Alexander_Carter} S. Alexander, J.S. Helman, and I. Balberg, Phys. Rev. B {\bf 13}, 304 (1976); 
S.A. Carter {\it et al.}, Phys. Rev. Lett. {\bf 67}, 3440 (1991).

\bibitem{Campbell} I. A. Campbell, A. D. Caplin and C. Rizzuto, Phys. Rev. Lett. {\bf 26}, 239 (1971); D. L. Mills, Phys. Rev. Lett. {\bf 26}, 242 (1971).

\bibitem{electronbook} P. Coleman in {\it Electron}, ed. by M. Springford (Cambridge University Press, Cambridge, 1997), Chap. 7.

\bibitem{Stewart} G. R. Stewart, Rev. Mod. Phys. {\bf 73}, 797 (2001).

\bibitem{Fawcett94} E. Fawcett, Rev. Mod. Phys. {\bf 66}, 25 (1994).

\bibitem{Hayden} S. M. Hayden {\it et al.}, Phys. Rev. Lett. {\bf 84}, 999 (2000).

\bibitem{Ong} T. R. Chien, Z. Z. Wang and N. P. Ong, Phys. Rev. Lett. {\bf 67}, 2088 (1991). 

\bibitem{Arajs} S. Arajs, Can. J. Phys. {\bf 47}, 1005 (1969).

\bibitem{Trego} A. L. Trego and A. R. Mackintosh, Phys. Rev. {\bf 166}, 495 (1968).

\bibitem{Komura} S. Komura, Y. Hamaguchi, and N. Kunitomi, J. Phys. Soc. Jpn. {\bf 23}, 171 (1967); Phys. Lett. {\bf 24A}, 299 (1967).

\bibitem{Barnes} R. G. Barnes and T. P. Graham, J. Appl. Phys. {\bf 36}, 938 (1965).

\bibitem{White}  G. K. White, R. B. Roberts and E. Fawcett, J. Phys. F {\bf 16}, 449 (1986).

\bibitem{McWhan} D. B. McWhan and T. M. Rice, Phys. Rev. Lett. {\bf 19}, 846 (1967).

\bibitem{Rice69} T. M. Rice {\it et al.}, J. Appl. Phys. {\bf 40}, 1337 (1969).
\bibitem{Bogenberger} B. Bogenberger and H. v. L\"{o}hneysen, Phys. Rev. Lett. {\bf 74}, 1016 (1995).

\bibitem {Tom}The increased sensitivity derives from both a factor of three smaller step size in $P$ at the QCP as well as a different constitutive relationship between $x$ and $P$ for CeCu$_{6-x}$Au$_{x}$ and Cr$_{1-x}$V$_{x}$.

\bibitem{Fedders} P. A. Fedders and P. C. Martin, Phys. Rev. {\bf 143}, 245 (1966).

\bibitem{Hertz_Milis} J. A. Hertz, Phys. Rev. B {\bf 14}, 1165 (1976); A. J. Millis, Phys. Rev. B {\bf 48}, 7183 (1993).

\bibitem{Chakravarty02} S. Chakravarty  {\it et al.}, Phys. Rev. Lett. {\bf 89}, 277003 (2002).

\bibitem{Norman}M. R. Norman {\it et al.}, Phys. Rev. Lett. {\bf 90}, 116601 (2003).

\bibitem{Hlubina} R. Hlubina and T.M. Rice, Phys. Rev. B {\bf 51}, 9253 (1995).

\bibitem{Schroeder} A. Schroeder {\it et al.}, Nature {\bf 407}, 351 (2000); 
P. Coleman {\it et al.}, J. Phys.: Cond. Matter {\bf 13},  R723 (2001).


\end{thebibliography}
\end{document}